\documentclass{aa}
\usepackage[english]{babel}
\usepackage[latin1]{inputenc}
\usepackage{graphicx}
\usepackage{aalongtable}
\usepackage{epsfig}

\usepackage{txfonts}
\newcommand{\vrad}{V$_{\rm rad}$}
\newcommand{\teff}{T$_{\rm eff}$}

\newcommand{\vmi}{V$-$I}
\newcommand{\vmic}{V$-$I$_{\rm C}$}
\newcommand{\vmik}{V$-$I$_{\rm K}$}

\newcommand{\kms}{\mbox{\rm km\,s$^{-1}$}}
\newcommand{\im}{I$_{\rm m}$}
\newcommand{\ic}{I$_{\rm C}$}
\newcommand{\imz}{I$_{\rm m}-z$}
\newcommand{\ks}{K$_{\rm s}$}
\newcommand{\imk}{I$_{\rm C}-$\ks}

\def\simgr{\,\hbox{\hbox{$ > $}\kern -0.8em \lower 1.0ex\hbox{$\sim$}}\,}
\def\simle{\,\hbox{\hbox{$ < $}\kern -0.8em \lower 1.0ex\hbox{$\sim$}}\,}

\begin{document}
\title{Detection of the lithium depletion boundary in the young open cluster  
IC~4665
\thanks{Based on observations collected at ESO-VLT, Paranal Observatory, 
Chile. Program number 073.D-0587(A).}}

%\subtitle{}

\author{S. Manzi\inst{1} \and S.Randich\inst{1} \and W.J. de Wit\inst{2,3} \and F. Palla\inst{1}}

\offprints{S. Randich, email:randich@arcetri.astro.it} %chiedere a cosa serve il comando offprints

\institute{INAF-Osservatorio Astrofisico di Arcetri, Largo E. Fermi 5, I-50125 Firenze, Italy 
\and 
Laboratoire d'Astrophysique, Observatoire de Grenoble, BP 53, F-38041 Grenoble, C\'{e}dex 9, France 
\and
School of Physics \& Astronomy, University of Leeds, Woodhouse Lane, Leeds LS2 9JT, UK}

\titlerunning{LDB in IC\,4665} 
\date{Received Date: Accepted Date}

% \abstract{}{}{}{}{}
% 5 {} token are mandatory

\abstract
{
The so-called lithium depletion
boundary (LDB) provides a secure and independent tool for deriving
the ages of young open clusters.
}
{
In this context, our goal is to determine membership for a sample
of 147 photometrically selected candidates of the young
open cluster IC\,4665 and to use confirmed members
to establish an age based on the LDB.
}
{
Employing the FLAMES multi-object spectrograph on VLT/UT2, we have
obtained intermediate-resolution
spectra of the cluster candidates. The spectra were
used to measure radial velocities and to infer the presence of the
\ion{Li}{i} 670.8~nm~doublet and H$\alpha$ emission.
}
{
We have identified 39 bona fide cluster members based on radial
velocity, H$\alpha$ emission, and Li absorption. The  mean radial
velocity of IC\,4665 is found to be \vrad$=\rm -15.95 \pm 1.13$\,km/s.
Confirmed cluster members display a sharp transition in
magnitude between
stars with and without lithium, both in the I$_{\rm m}$ vs.~\imz~and
in the K$_{\rm s}$ vs. \imk~diagrams. From this boundary, we
deduce a cluster age of
$27.7^{+4.2}_{-3.5} \pm 1.1 \pm 2$~Myr.
}
{
IC\,4665 is the fifth cluster for which an LDB age has been
determined, and it is the youngest cluster among these five. Thus, the LDB is
established from relatively bright stars still in
the contracting pre-main sequence phase. The mass of the boundary
is M$_\ast=0.24 \pm 0.04$~$M_\odot$. The LDB age agrees well with the
ages derived from isochrone fitting of both low and high mass, turn--off
stars, a result similar to
what is found in the slightly older NGC~2547.
}

\authorrunning{S. Manzi}
\maketitle
\section{Introduction}\label{intro}
Pre-main sequence (PMS) clusters with an age of 5-50 Myr represent an
ideal tool for investigating several aspects related to star formation
and the early phases of (sub-)stellar evolution. 
These clusters provide the youngest samples of PMS stars
outside a star forming environment.
Indeed, unlike star forming
regions, PMS clusters show the complete and final product of the
formation process, immediately after the active phase of star
birth. Moreover, the low-mass members (both stars and brown dwarfs)
are still bright and readily detectable, and not
affected by severe extinction since most of the
circumstellar and interstellar material has been
accreted and dispersed. Finally, the age interval of PMS clusters
is critical with respect to the early 
evolution of protoplanetary disks, stellar
rotation, and activity.
Despite these benefits, at present only three
systems are confirmed PMS clusters: NGC~2547 ($\sim$35~Myr; Jeffries \&
Oliveira~\cite{jo05}), NGC~2169 ($\sim$10~Myr;
Jeffries et al.~\cite{jeff07}), 
IC~2391 ($\sim 50$~Myr; Barrado y Navascu\'es
et al.~\cite{byn}); a few other candidates exist.    
The identification of additional PMS cluster candidates and confirmation
of their nature through a 
secure age determination would therefore represent a significant improvement
from the phenomenological and statistical point of view
(see discussion in Jeffries et al.~\cite{jeff07}).

Ages of stellar clusters can be obtained from the location of
the main sequence turn-off 
(t$_{\rm nucl}$) or from the isochronal distribution
of the PMS population in the HR
diagram. These classical methods are widely used, but suffer from large
uncertainties of up to a factor of two in age 
(e.g., Mermilliod~\cite{merm00}; Jeffries \& Oliveira~\cite{jo05}). 
On the contrary, the method based on the
lithium depletion boundary (LDB) has proven to be robust and less model 
dependent since it relies on well known physics
(Bildsten et al.~\cite{bildsten}; Ushomirsky et al.~\cite{ush}). 
During the PMS phase, stars undergo
a gradual gravitational contraction that causes a progressive,
mass-dependent rise of the central temperature. Li burning starts when the core
reaches a temperature $\simeq 3\times 10^{6}$\,K 
(depending on density); hence, in fully convective low-mass stars 
(M $\la 0.5 \rm M_{\odot}$) Li is depleted from the initial 
interstellar abundance on
a time scale that is a sensitive function of mass. In a young stellar
cluster three regimes of Li depletion are present: i) relatively massive stars
(with radiative interiors) that suffer only a little amount 
of Li depletion; ii) stars in the so-called Li chasm (Basri~\cite{basri97})
that have fully depleted their initial Li supply; iii) low mass
stars that have 
preserved the initial Li content. The transition between low-mass
stars with and without Li is very sharp and the luminosity of 
the faintest star that has depleted 99 \% of its initial Li
identifies the boundary (LDB) and the age (t$_{\rm LDB}$) of 
the cluster (e.g., Basri et al.~\cite{bmg96}; Basri~\cite{basri97};
Stauffer~\cite{stauffer00}). The older the cluster is, the
fainter the stars at the boundary are. So far, the LDB has been detected
in four clusters: the Pleiades (Stauffer et al.~\cite{stauf98}), 
$\alpha$ Persei (Stauffer et al.~\cite{stauf99}),
IC\,2391 (Barrado y Navascu\'es et al.~\cite{byn}),
and NGC\,2547 (Jeffries \& Oliveira~\cite{jo05}). 
Remarkably, the LDB ages
determined for the Pleiades, $\alpha$ Per, and IC\,2391 exceed 
the nuclear ages by a factor $\sim 1.5$; these particular nuclear ages
were derived from fitting the cluster turn-off (TO)
with evolutionary models without overshooting.
On the other hand, the two dating methods yield similar ages
for the younger cluster NGC~2547, although the cluster's nuclear age
is still rather uncertain.

In this paper, we report on the determination of the LDB 
in the open cluster IC\,4665 whose properties have been 
extensively discussed by
de Wit et al.~(\cite{dewit}). 
IC\,4665 is an interesting candidate PMS cluster,
located relatively far from the Galactic plane at 
$b \sim +17^\circ$. Its nuclear age
is about t$_{\rm nucl}=36\,{\rm Myr}$ (Mermilliod~\cite{merm81}), 
but other properties might suggest
an age as high as 100~Myr (Prosser~\cite{pros93}). The Hipparcos
distance is $385\pm40$\,pc (Hoogerwerf et al. 2001), while a lower value of
320\,pc has been derived by Crawford \& Barnes (1972\nocite{}). IC\,4665 
was targeted for a wide and deep survey in I$_{\rm m}$ (Mould) and $z$ 
filters at the Canada-France-Hawaii telescope (CFHT). 
This deep photometric survey led to the
detection of 786 new low-mass stellar and brown dwarf candidate
members ($14.8< \rm I_{\rm m}<22$), down to about $30$M$_{\rm
jup}$ (de Wit et al.~\cite{dewit}). 
Given the nuclear age of IC\,4665, the new
low-mass candidate members provide a sample suitable for the detection 
of the LDB and for the derivation of an accurate and independent age estimate.
The importance of such determination was already stressed by Mart\'\i n 
\& Montes
(1997) who were the first to obtain Li abundances in a small sample
of cluster stars (mainly of G and early-K spectral-type).
Although a spread in Li was found,
the observations did not reach the low luminosity population
of IC~4665 where the LDB is expected to occur.

Our paper is structured as follows: Section 2 describes the sample selection, 
observations, and data analysis. The results on membership and the lithium 
boundary are given in Section 3. The age of IC~4665 and the comparison with 
other young PMS clusters is discussed in Section 4. The conclusions close 
the paper.
\begin{figure}
\includegraphics[width=9cm]{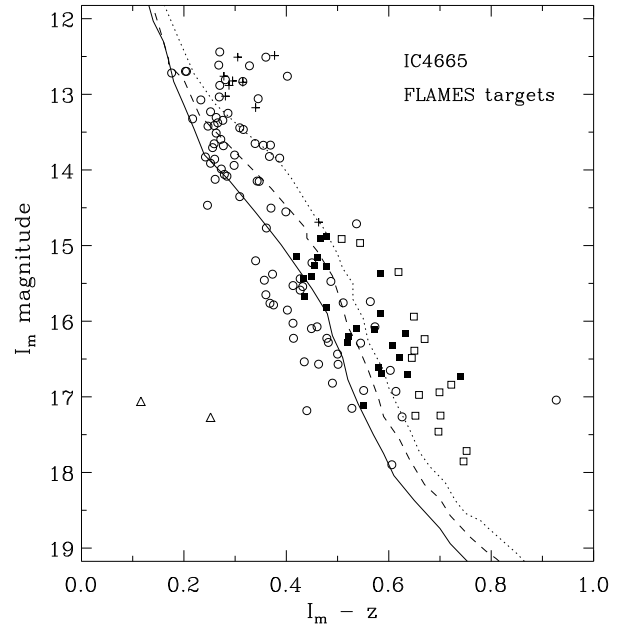}
\caption{I$_{\rm m}$ vs. I$_{\rm m}-z$ color-magnitude diagram
of all the 137 
candidate members observed with Giraffe and with available photometry 
from the CFHT survey.
Target stars have been retrieved from {\bf i.}
Prosser (1993) and Giampapa et al.~(\cite{giampapa}): open circles, 89 stars;
{\bf ii.} de Wit et al. (\cite{dewit}), open 
squares and crosses --
15 and eight stars, respectively; {\bf iii.}
Luyten (1961): triangles, two stars. 
Filled squares instead
represent candidates common to Prosser and de Wit et al. (24 stars).
Crosses are the eight objects brighter than the 
minimum magnitude selection limit in de Wit et al. and not reported in 
Prosser. Three isochrones for 30, 50, and 100~Myr are shown from the Nextgen 
model of Baraffe et al. (1998) with a mixing length parameter $\alpha=1$ and
for a distance of 370\,pc (see Sect.~\ref{boundary}).}
\label{cmd1}
\end{figure}
\section{Sample selection, observations and data analysis}
Target stars were selected from different sources: 
Prosser~(\cite{pros93}), Giampapa et al.~(\cite{giampapa}),
de Wit et al.~(\cite{dewit}), and a literature search
on SIMBAD. The sample includes 96 stars from Prosser (1993 --of
these, 88 were covered by the CFHT survey, but not included in the catalog
of de Wit et al.), one star from Giampapa et al.~(\cite{giampapa}),
15 stars from 
de Wit et al. (2006), 24 stars included both in Prosser and de Wit et al.
catalogs, and eight new candidate members.
The latter had been selected by de Wit et al., but were
not included in their final analysis since they are brighter than 
I$_{\rm m}=$14.8~mag, the cutoff limit used in that paper. 
Finally, we also observed two stars from Luyten (1961) and one 
IRAS source (IRAS 17447$+$054), since they 
happen to lie in the FLAMES target fields. The present 
sample hence consists of a total of 147 stars. Of these, a subsample of 137 
candidate cluster members have I$_{\rm m}$ and $z$ photometry available 
from the CFHT survey. Fig.\,\ref{cmd1} shows their distribution in the 
color-magnitude diagram. Stars with I$_{\rm m}$ magnitudes between $\sim 16$
and 18 bracket the expected apparent brightness of the LDB
boundary at the distance of IC~4665 corresponding 
to a relatively young ($\sim$~20\,Myr) and old ($\sim$~50\,Myr) age. 
\begin{figure*}
\includegraphics[width=8.5cm]{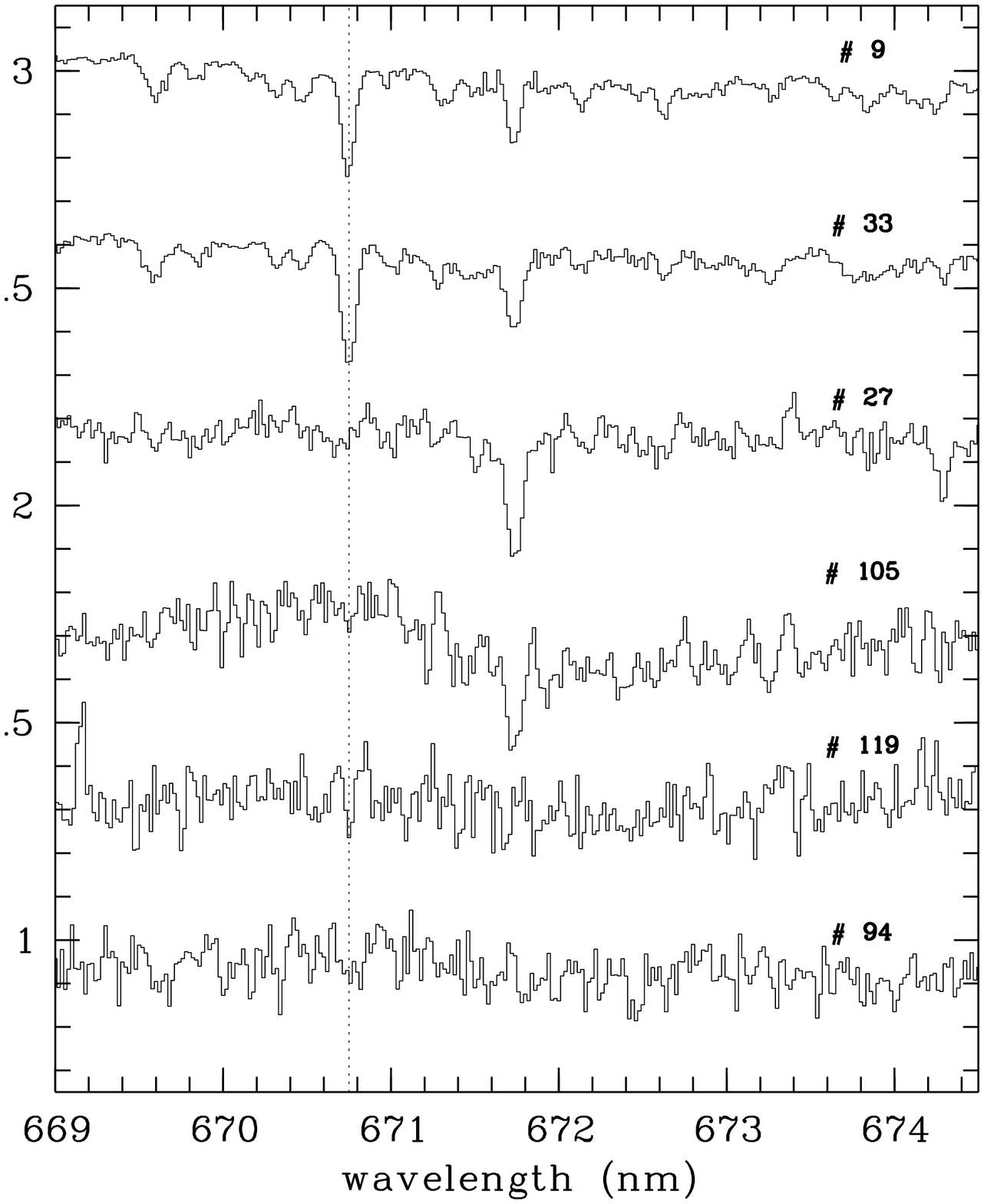}
\includegraphics[width=8.5cm]{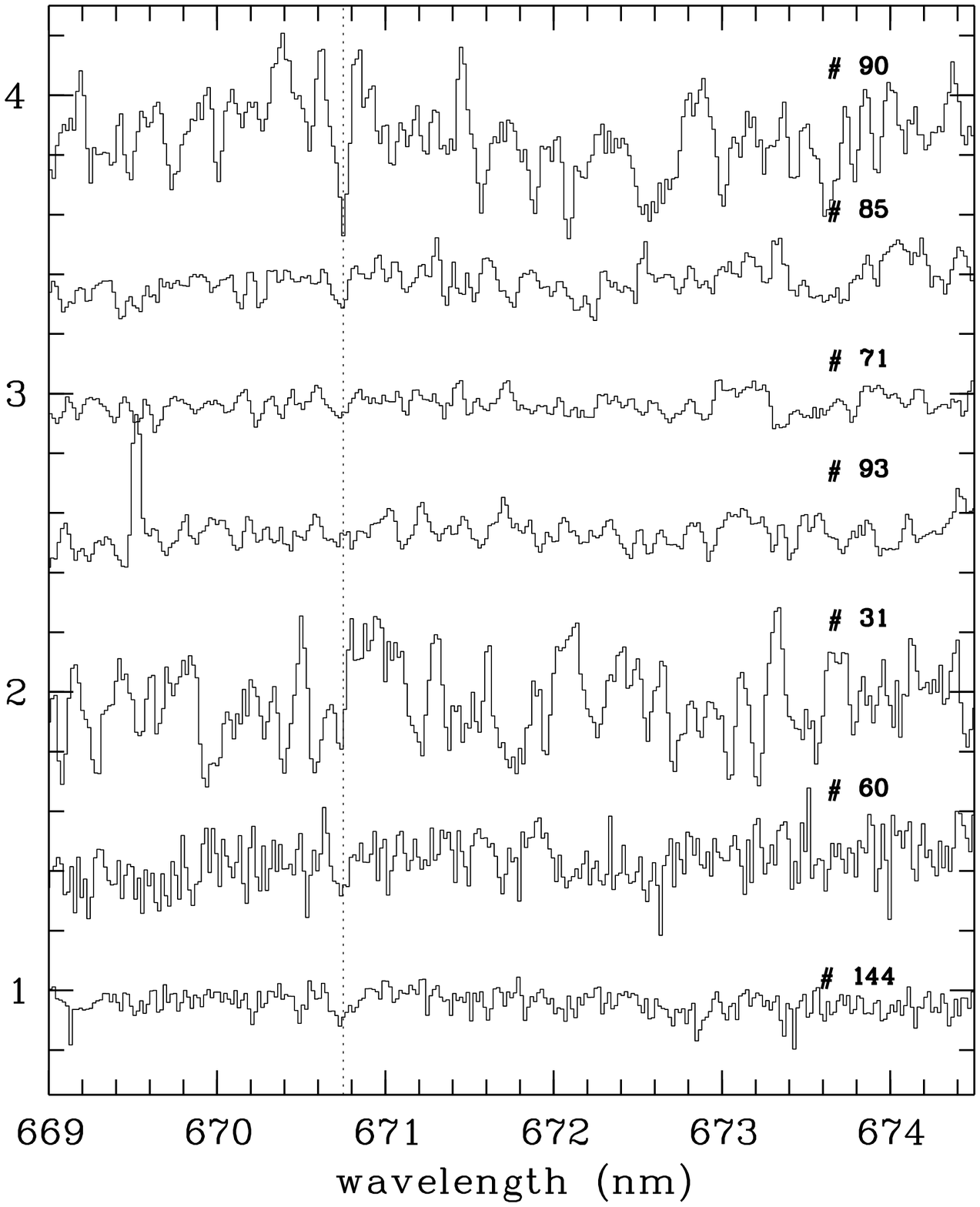}
\caption{Spectra of 13 stars ordered by increasing magnitude representative 
of the Li range: above the chasm (\#9 and \#33), 
in the chasm (\#27 and \#105), and
close and below the the LDB
(\#119, \#94, \#90, \#85, \#71, \#93, \#31, \#60, \#144).
The vertical dashed 
line at 607.8~nm marks the expected Li line position. Note that the TiO
bands at wavelengths redder than $\sim$~672~nm are not visible in the
spectra with very low S/N due to poor sky subtraction.
} 
\label{spectra}
\end{figure*}
The 147 target stars are listed 
in Table~\ref{table1} 
that gives a running number (Col. 1); coordinates (Cols. 2-3); 
data source (Col. 4: 1- only in Prosser or Giampapa et al.; 
2- from Prosser and de Wit et al.;
3- only in de Wit et al.; 4- not in Prosser nor in de Wit et al. because
brighter than the I$_{\rm m}$ limit of that paper; 9- from IRAS or Luyten); 
name (Col. 5); I$_{\rm m}$ 
and $z$ magnitudes (Col. 6-7); radial velocity and associated 
error (Col. 8-9; a ``-" means that we were not able to measure the
velocity); presence (Y) or lack (N) of the Li line (Col. 10:
``?" means that the S/N did not allow to confirm or reject the presence of Li); 
presence of the H$\alpha$ 
line in absorption or emission (Col. 11); and membership flag (Col. 12 --see
Sect.~3.1).
The I$_{\rm m}$ and $z$ magnitudes listed in the table
are from the CFHT measurements. Note that for a large fraction
of these objects VI photometry in the Kron system is also available
from Prosser (1993) and that near-infrared photometry for all the stars 
was retrieved from the Two Micron All Sky Survey (2MASS) catalogue.
In the following, we will use our running numbers given in Col.~1.

The observations were carried out in Service Mode during
May, June, July, and September 2004 using the FLAMES instrument (Pasquini
et al. 2002) on VLT/UT2. The spectra were
obtained with the GIRAFFE spectrograph in conjunction with the MEDUSA
fibre system and a 600 lines/mm grating (L6). The resolving power is
R~$\sim 8000$, and the exposure time chosen to reach a S/N ratio
$\sim$~20 for the faintest targets. The spectra cover the
wavelength range between 643.8\,nm and 718.4\,nm, which includes the H$\alpha$ 
line, besides the Li~{\sc i} 607.8~nm resonance doublet. 
The stars were observed in two configurations, A and B,
covering two different fields. The configurations were centered at
RA(2000)=17h~46m~46.27s and DEC(2000)=$+$05d~38m~17.7s, and
RA(2000)=17h~45m~26.91s and DEC(2000)=$+$05d~56m~53.7s, respectively.
94 and 53 stars were allocated in configurations A and B.
We obtained six 45~min exposures for configuration A
and four 45~min exposures for configuration B. 

Data reduction was performed using
Giraffe BLDRS\footnote{version 1.12 -- http://girbldrs.sourceforge.net/},
following the standard procedure and steps (Blecha \& Simond 2004).
The sky contribution was subtracted separately; namely, for each
configuration, we considered 15 sky spectra, subdivided them in three groups
of five spectra, and derived the median sky from each group. 
Then, we formed a ``master" sky 
by taking the average of the three median sky spectra.
Due to the fact that the sky on the CCD was
rather inhomogeneous and to the presence of scattered light from the fibers
allocated to very bright objects, for the faint stars
an appropriate sky subtraction was impossible to perform. 
For this reason, while both the Li and H$\alpha$ lines when present are 
in most cases
clearly visible in the spectrum, we prefer not to give any quantitative 
measurement of their equivalent widths.

Data handling and analysis has
been carried out both with MIDAS and IRAF\footnote{
IRAF is distributed by the National Optical
Astronomical Observatories, which  are operated by the Association of
Universities for Research in Astronomy,  under contract with the National
Science Foundation.}
software packages. In most cases multiple sky-subtracted spectra of the same
target have been combined, after adjusting them for Doppler shift due to
the motion of the earth after sky subtraction. 
For a few stars we excluded one or more exposures, due to 
bad quality. Final S/N ratios per resolution element
are in the range $\sim 200-15$.
In Fig.\,\ref{spectra} we present some representative spectra spanning the
magnitude range corresponding to the three regimes mentioned in 
Sect.~\ref{intro}: solar-type stars
with a strong Li line; stars lacking Li that fall
in the Li chasm; stars below the LDB, showing again the Li feature.

Radial velocities (\vrad) have been measured from the average
shift of the spectral lines in the co-added spectra.  For some
critical spectra (e.g. low S/N, suspected binary) we have determined
\vrad~from the individual exposures.
Measurements of \vrad~were carried out using IRAF
and the $RVIDLINES$ procedure. We typically
used 10-20 lines per star, depending on S/N.
For the faintest stars, \vrad~was determined using a couple of lines only.
Resulting heliocentric radial velocities have errors between 0.5 and 6~km/sec
and are listed in Cols. 8 and 9 of Table~\ref{table1}.
\setcounter{table}{1}
\begin{table*}
\caption{Stars confirmed as members. The values for I$_{\rm c}$ are converted
from I$_{\rm m}$, as explained in 
the text.}\label{table2}
      \begin{tabular}{rcccc|rcccc|rcccc}
        \hline
 \#  & I$_{\rm m}$  & I$_{\rm m}-z$ & I$_{\rm c}-K_{\rm s}$ & EW(Li)  &  \#  & I$_{\rm m}$  & I$_{\rm m}-z$
 & I$_{\rm c}-K_{\rm
s}$ & EW(Li) & \#  & I$_{\rm m}$  & I$_{\rm m}-z$ & I$_{\rm c}-K_{\rm s}$ &  EW(Li) \\
     &  (mag)  & (mag) & (mag) & (m\AA) &  & (mag) & (mag)& (mag) & (m\AA) & & (mag) & (mag) & (mag) & (m\AA) \\
 &  &   &  &  &  &  &  & & &&  &   &  &    \\
   \hline
 20 & 12.440 & 0.270 & 1.639 &  30$\pm 4$  & 	  72 & 14.876 & 0.478 & 2.449 & --- &	 100 & 16.475 & 0.621 & 2.709 & --- \\
  6 & 12.614 & 0.268 & 1.771 &  52$\pm 5$  &	  88 & 14.910 & 0.467 & 2.287 & --- &	  69 &  ---   &  ---  &  ---  & ---\\
  9 & 12.623 & 0.328 & 1.343 & 235$\pm 5$  &	  95 & 15.228 & 0.450 & 2.482 & --- &	 119 & 16.608 & 0.580 & 2.506 & Y? \\ 
 26 & 12.696 & 0.203 & 1.243 & 255$\pm 7$  &	 147 & 15.269 & 0.454 & 2.302 & --- &	 94 & 16.650 & 0.603 & 2.710 & Y? \\
 33 & 12.697 & 0.205 & 1.291 & 274$\pm 7$  &	 105 & 15.278 & 0.477 & 2.369 & --- &	 90 & 16.685 & 0.585 & 2.665 & Y \\ 
 74 & 12.718 & 0.176 & 1.255 & 245$\pm 10$ &	 145 & 15.373 & 0.584 & 2.591 & --- &	 85 & 16.706 & 0.637 & 2.661 & Y \\ 
 62 & 13.074 & 0.233 & 1.567 & 236$\pm 8$  &	  55 & 15.474 & 0.487 & 2.412 & --- &	 71 & 16.736 & 0.741 & 2.965 & Y \\ 
 16 & 13.413 & 0.259 & 1.569 &  35$\pm 4$  &	  98 & 15.741 & 0.564 & 2.590 & --- &	  93 & 16.840 & 0.722 & 2.730 & ? \\  
 59 & 13.443 & 0.309 & 1.959 &  70$\pm 3$  &	  87 & 15.902 & 0.584 & 2.545 & --- &	 31 & 16.928 & 0.614 & 2.747 & ? \\
125 & 13.680 & 0.277 & 1.699 &  31$\pm 4$  &	  82 & 15.940 & 0.649 & 2.579 & --- &	 60 & 16.940 & 0.699 & 2.860 & Y \\
 83 & 13.705 & 0.256 & 1.748 &   ---       &	  28 & 16.074 & 0.573 & 2.484 & --- &	 144 & 17.248 & 0.701 & 2.896 & Y \\
142 & 13.844 & 0.387 & 2.464 &   ---       &	 128 & 16.116 & 0.572 & 2.550 & --- &	      &  &  &  &   \\
 27 & 14.147 & 0.343 & 1.959 &   ---       &	 121 & 16.169 & 0.632 & 2.720 & --- &	     &  &  &  &   \\
120 & 14.554 & 0.399 & 2.262 &   ---       &	 138 & 16.236 & 0.670 & 2.814 & --- &        &  &  &  & \\	
        \hline
\end{tabular}
\end{table*}
\section{Results}
\subsection{Membership}\label{memb} 
\begin{figure}
\includegraphics[width=9cm]{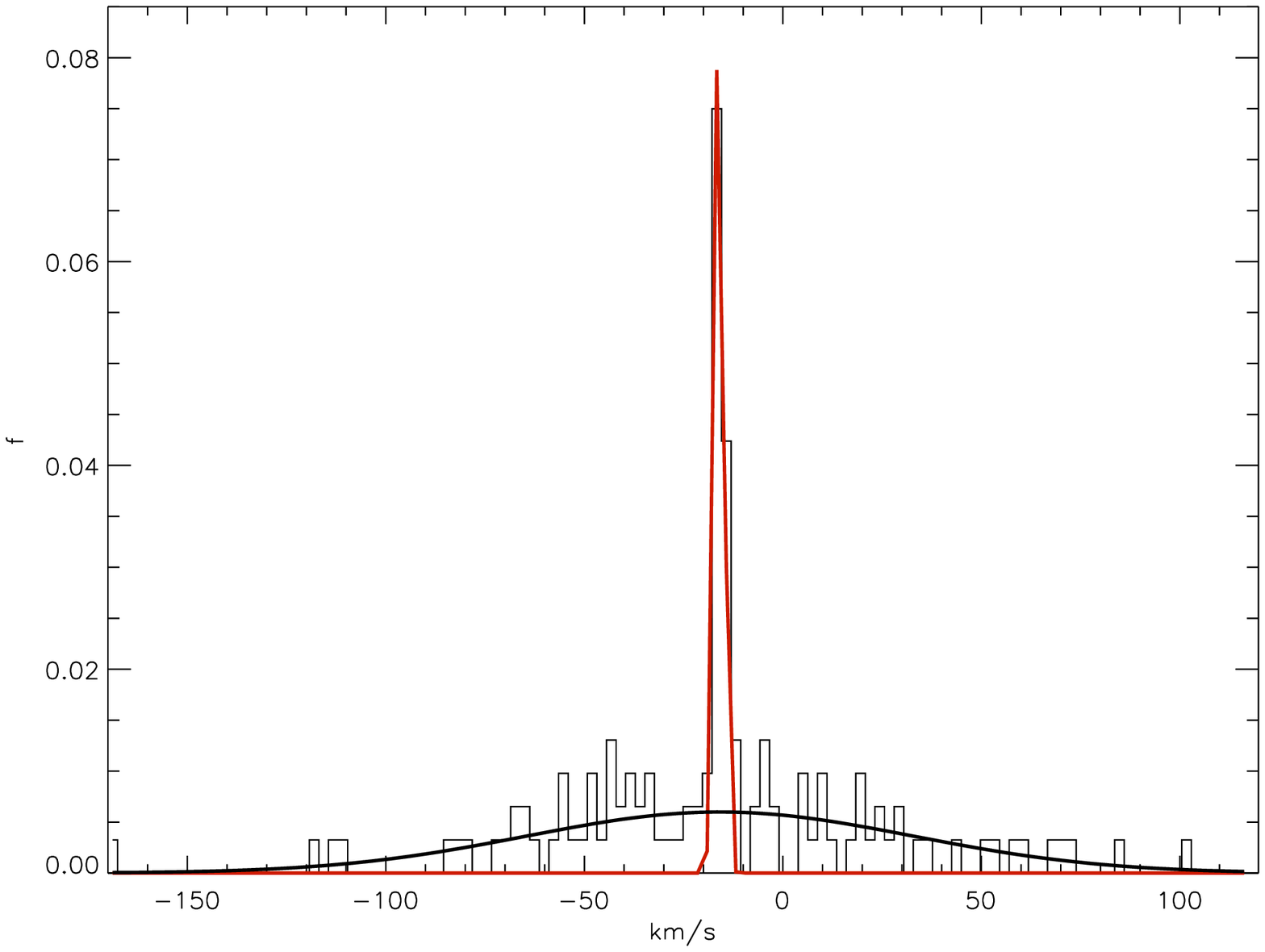}
\caption{Radial velocity distribution of IC~4665 candidate members.
The two gaussians (solid curves) indicate the best fits for the cluster and 
field, respectively.}  \label{vrad}
\end{figure}

In order to confirm or reject membership of the cluster candidates,
we applied the usual radial velocity criterion, together with the requirement 
on the presence of Li (for bright stars) and/or H$\alpha$.
We have first estimated the cluster average \vrad~ and its
standard deviation from the observed \vrad~ distribution of the
sample. In Fig.~\ref{vrad} we show the distribution of
measured radial velocities 
(stars with variable \vrad~are obviously not included in the figure): 
there is a clear peak at \vrad $\approx -$15 km/s which indicates the presence 
of the cluster. 
The average velocity was derived by fitting the observed distributions
with two gaussian functions, one for the cluster and one for field stars;
the best fit was then determined using a maximum likelihood algorithm. 
For IC~4665 we find
\vrad (IC~4665)$=-15.95$~km/s and $\sigma$(IC~4665)$=$1.13~km/s, while for 
the field we obtain 
\vrad~(field)$=-$15.87~km/s and $\sigma$(field)$=$48.63~km/s. 
Interestingly, the field and cluster velocities
are very similar, although, as expected, the distribution of the field stars
is much broader. The cluster average velocity is
compatible with previous estimates, notably that from
the high-mass members {\it viz.}
\vrad$=-15.5$~km/s and $\sigma=2.9$~km/s 
(Crampton et al. \cite{cram76}). On the other hand, Prosser \& Giampapa 
(\cite{pg94}) found the slightly higher value of \vrad$=-13$~km/s.
We have considered as cluster
members those stars with \vrad~within 
$\pm 3\sigma$ from the average value.
In addition, we have also included three stars (\#55, \#87, 
and \#138) 
with \vrad~slightly outside this limit, but
with large errors on \vrad~and with other indicators consistent
with membership (see below). With this criterion 42 radial velocity
members were found, with an expected
statistical contamination (i.e., non
members with \vrad~consistent with the cluster) of five stars, as
estimated from the fitting procedure.
Furthermore, 14 possible members with variable radial velocity
and/or evidence for a double line system were considered as possible members.
Thus, we have a total of 56 possible candidates.
\begin{figure}
\psfig{figure=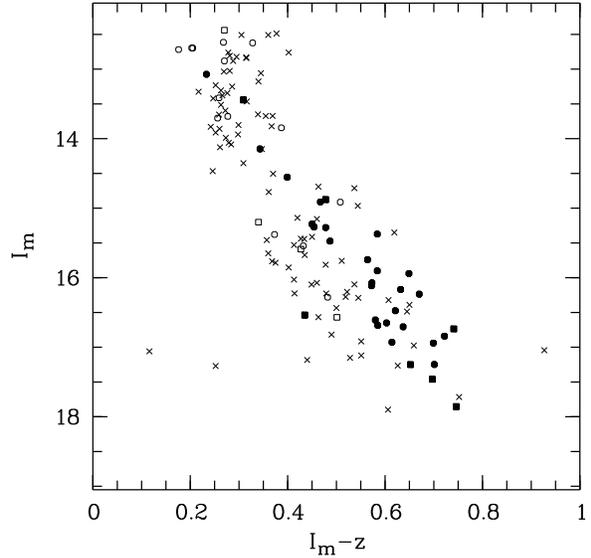,height=8cm, angle=-90}
\caption{CM diagram of the observed targets with available Iz photometry
including information on \vrad~and
presence of H$\alpha$ emission. Symbols are as follows:
open circles - \vrad~members without H$\alpha$ emission;
filled circles - \vrad~members with H$\alpha$ emission, including
the three marginal \vrad~members;
open squares - \vrad~variables without H$\alpha$ emission;
filled squares - \vrad~variables with H$\alpha$ emission or "abs/em";
crosses - all other stars.}\label{halpha}
\end{figure}

The presence of H$\alpha$ emission 
and/or Li absorption provides additional membership
criteria for the \vrad~ members and for stars without a
radial velocity measurement. In Fig.~\ref{halpha} we
show  the same color-magnitude (CM)
diagram of  Fig.~\ref{cmd1}, but with the additional
information on \vrad~and H${\alpha}$ of each individual star. The 
figure clearly shows that
stars with I$_{\rm m}\ga$14 must have H$\alpha$ emission 
in order to be members. Therefore, we considered as non members all stars
fainter than this magnitude limit without H$\alpha$ emission: we found 10
such stars (\#2, \#36, \#41, \#44, \#48, \#49, \#67, \#81, \#126, \#139), 
four of them with
\vrad~ consistent with membership, and six with variable \vrad. 
We also discarded the star \#129 which does not show the Li line,
in spite of having a magnitude brighter than the Li chasm, as well as stars \#5
and \#141, with no available photometry, but lacking both Li and H$\alpha$.
Interestingly, the number of stars with consistent \vrad, but not satisfying
the H$\alpha$/Li criteria (seven) is comparable to
that of the expected number of contaminants.
Finally, we note the presence of four stars with uncertain 
membership status,
because we could not retrieve 
enough/secure information from their spectra. One of them, star \#11,
is a SB2 system with H$\alpha$ emission and uncertain Li status; given
its position in the CM diagram (below the sequence), it is likely a non 
member.
A final membership flag is given in Col.~12 of Table~\ref{table1}.
Specifically, we assigned a ``Y'' status to stars with secure membership and
consistency between all indicators; 
we marked ``Y?' stars with two indicators
(out of three) consistent with membership and the three stars with \vrad~
slightly outside the permitted range, while
we gave uncertain status
(``?'') to the four stars with variable radial velocity, 
H$\alpha$ emission,
and uncertain Li line. Finally, stars that turned out to be non members
are marked with ``N".
In summary, of the 147 candidates, 39 stars are members (Y or Y? status),
104 non members, and 4 with uncertain membership.
The 39 confirmed members are listed in Table~\ref{table2}, where we also give 
the Li equivalent widths for the brighter objects and the presence of the 
Li line for the faintest ones.

Our analysis indicates that 27\% of the sample stars are likely cluster 
members, in reasonable agreement with the estimate of de Wit et al.~(\cite
{dewit}). A detailed discussion of membership and contamination
in different mass bins cannot be presented here, since not only our
spectroscopic sample is incomplete as a whole, but stars in different mass bins
are characterized by varying degrees of completeness.
This discussion is deferred to a forthcoming paper
(de Wit et al. in preparation) where the analysis
of a much larger sample of very low-mass cluster stars and brown dwarfs will
be presented, based on low resolution optical and near-IR spectra.
\subsection{The absolute brightness of the Li boundary}\label{boundary}
\begin{figure*}
\includegraphics[width=8cm]{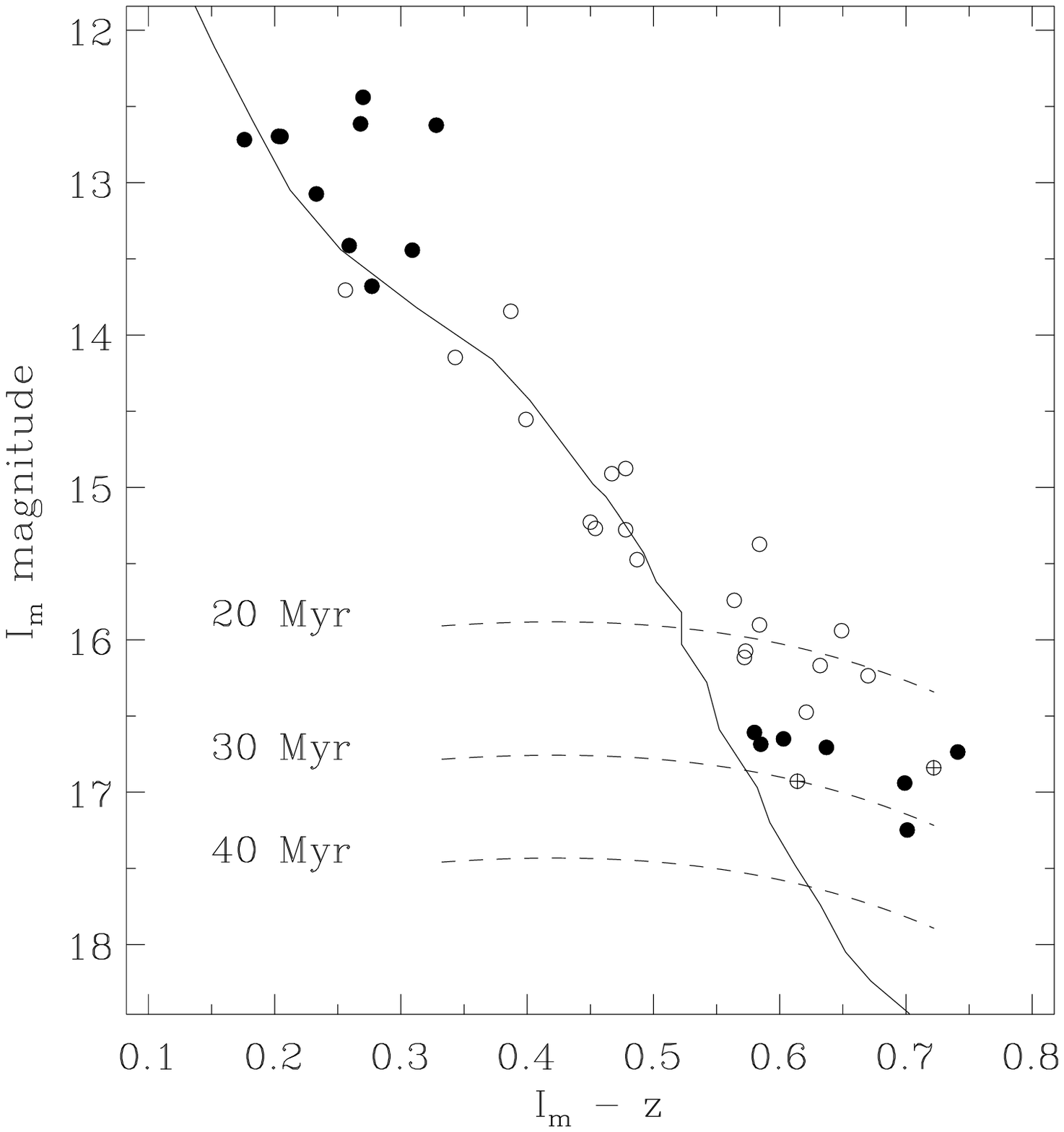}
\includegraphics[width=8cm]{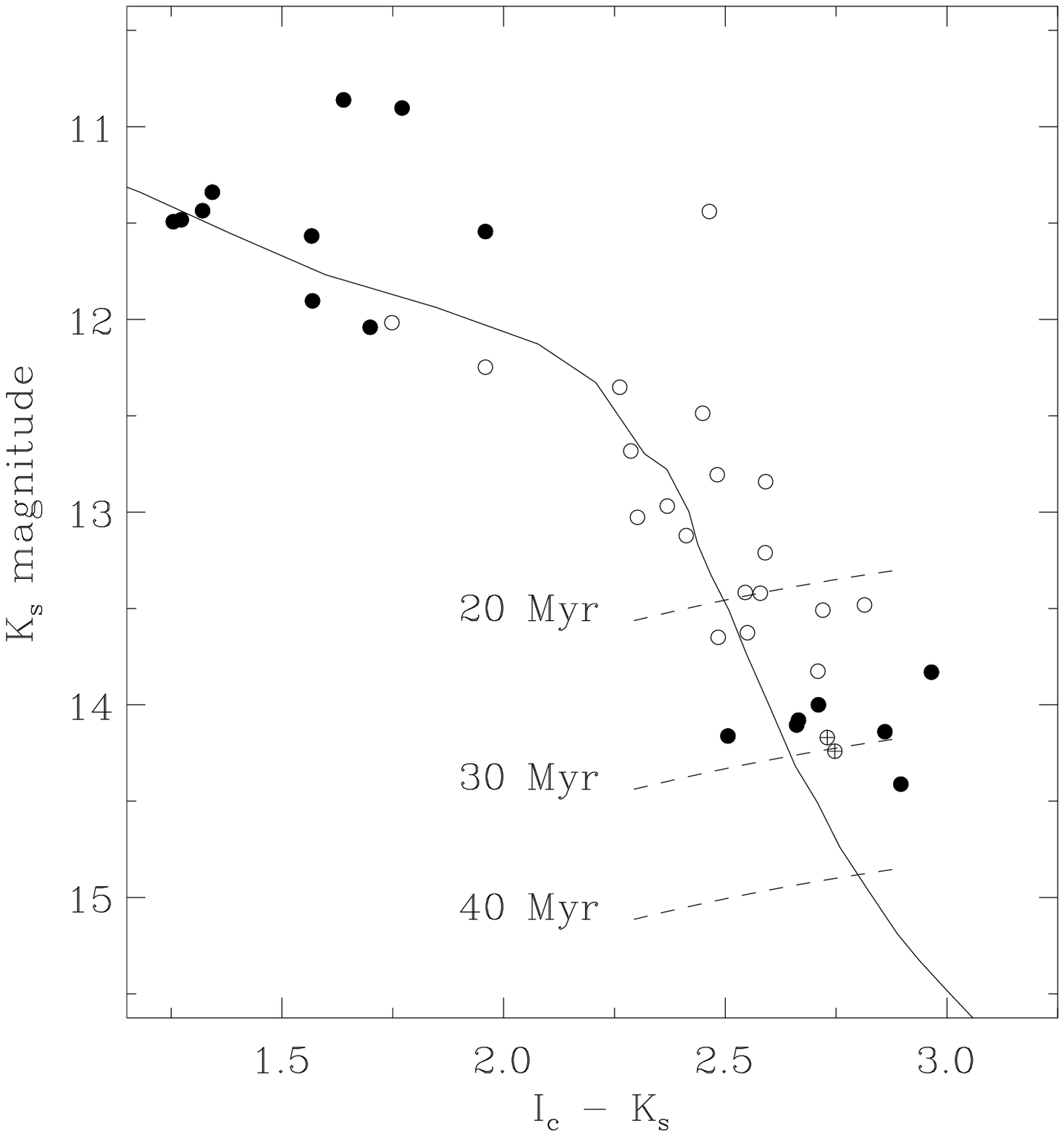}
\caption{I$_{\rm m}$ vs. I$_{\rm m}-z$ (left panel) and K$_{\rm s}$ vs. 
I$_{\rm C}-\rm K_{\rm s}$ (right panel)
diagrams of the 38 likely cluster members with available photometry
listed in Table~\ref{table2}. Filled and open symbols indicate stars
with detected/undetected Li line, respectively. The two crossed circles 
denote stars with uncertain Li detection. The theoretical
position of the boundary from Chabrier \& Baraffe~(\cite{cb97}) models 
for ages of 20, 30, and 40 Myr is also shown (see text), along with 
the 30~Myr isochrone. 
Note that, by definition, the LDB is actually
just a point on the isochrone. What we show here,
following Jeffries \& Oliveira (\cite{jo05}), are constant luminosity
curves, with the luminosity corresponding to the LDB luminosity
at a given age. These curves are not flat since, for a given M$_{\rm bol}$,
the bolometric correction changes with color (\teff). To construct
the curves, we estimated bolometric corrections for
\im~and \ks~as function of color from Baraffe et al.~(\cite{bar98}) models.
}\label{cm_ldb}
\end{figure*}

In Fig.~\ref{cm_ldb} we show the \im~vs. \imz~and \ks~vs. \imk~color-magnitude 
diagrams of the 38 likely cluster members listed in Table~\ref{table2} with
available photometry.

A chasm and boundary are clearly present in both diagrams. We determine 
the observed LDB of IC\,4665 from the brightest star with secure
Li detection (``Y" status in Table~\ref{table2}) 
on the faint side of the chasm. 
In the \im~vs. \imz~diagram 
this corresponds to the star \#90 with  I$_{\rm m}=16.68$. 
Note that, whereas the Li line might be present in the spectra of stars \#119
and \#94 (classified as ``Y?" in Table~\ref{table2}), the lower S/N of these
two spectra makes the detection of Li less secure than in the case of star
\#90.
Alternatively, the theoretical LBD could also be defined by the
faintest star within the chasm.  In such a case, the LDB of IC\,4665 would 
corresponds to the star \#100 with I$_{\rm m}=16.47$~mag 

For most of faint members in this paper, standard
VI$_{\rm C}$ photometry is not available. In order to
convert Mould to Cousin I magnitudes, we used
new photometry of IC\,4665
that will be presented in James et al. (\cite{james}). More specifically,
James et al. have
performed a shallow survey of IC\,4665 in BVI$_{\rm C}$, allowing 
us to derive relationships both
between \imz~and \vmic~colors and between \ic~and \im~
magnitudes. The relation between \imz~and \vmic~was obtained
for objects present in both James et al. and de Wit et al. (\cite{dewit})
with $0.0<$\imz$<0.35$
or, correspondingly, 0.7$<$V--I$_{\rm C}<$2.1. We found
\vmic=0.65444+3$\times$(\imz)+3$\times$(\imz)$^2$. This was then extrapolated
up to \imz$=0.75$. Using the 1$\sigma$ errors on the fit 
and extrapolating the 1$\sigma$ upper and lower limit
of the fit, delivers a range in values for V--I$_{\rm C}$ of 0.3~mag.
As to magnitude conversion, we directly compared \im~and \ic~magnitudes of
stars in common in the two surveys and found median values
\ic$-$\im$=$0.03 and 0.06
for \imz~lower and greater than 0.2~mag, respectively. The typical scatter is 
0.05~mag. 

As a check, {\bf 1)} we estimated the expected \ic$-$\im difference for
stars of different temperatures based on
the filter transmission curves and found that it increases from $\sim 0.03$~mag
at 6000~K to $\sim 0.06$~mag at 3500~K; {\bf 2)}
we derived \vmic~colors from \vmik~colors
published by Prosser~(\cite{pros93}), employing the transformation
of Bessel~(\cite{bes79}); from these \vmic~colors and V magnitudes given in
Prosser, we also estimated \ic~magnitudes.
We found a good agreement between magnitudes and colors estimated in
this way and those
obtained extrapolating from James photometry: namely,
$\Delta$(\vmic)$_{\rm mean}=0.02 \pm 0.1$~mag and
$\Delta$\ic$_{\rm mean}=0.1 \pm 0.28$~mag.

As mentioned in Sect.~\ref{intro},
the distance to IC\,4665 is not accurately known with
values ranging from 320\,pc (Crawford \&
Barnes 1972\nocite{}) to 385+40=425\,pc (Hoogerwerf et
al. 2001\nocite{}). Without convincing evidence for a short or 
a long distance, we adopt a compromise between the two extremes,
i.e. a distance of 370$\pm$50\,pc. We have also estimated our own 
distance to IC\,4665
by comparing the photometry (V, B--V) of high mass stars in IC\,4665 to
the Pleiades and determining a vertical offset. By assuming a distance
to the Pleiades of 133~pc and E(B--V)=0.03, the best matching
of the two sequences is obtained with a distance to IC\,4665 equal to 366~pc,
very close to our adopted average.

Taking A(\ic)=0.333 (from E(B$-$V)$=$0.18 
--Hogg \& Kron~\cite{hg55}- and the extinction law of 
Dean et al.~\cite{dean}), we obtain for stars \#100 and \#90
M$_{\rm I_{c}}=8.37$ and 8.57, respectively.
Considering the average of the 
two values, we find that the LDB occurs at 
M$_{\rm I_{c}}=8.47^{+0.32}_{-0.28}\pm 0.10$~mag, where the first 
contribution
to the error is due to uncertainty in distance and the second one
reflects the uncertainty in the LDB determination. Clearly the
error is dominated 
by the distance uncertainty. 

Similarly, the brightest/faintest stars with/without Li in the \ks~vs. \imk~
diagram are \#90 (\ks=14.08) and \#100 (\ks=13.826).
Assuming A$_{\rm K}=0.06$~mag, this yields an LDB at 
\ks=$6.05^{+0.32}_{-0.28}\pm 0.13$. 
\subsection{The LDB age of IC\,4665}
\begin{figure*}
\includegraphics[angle=-90,width=8.5cm]{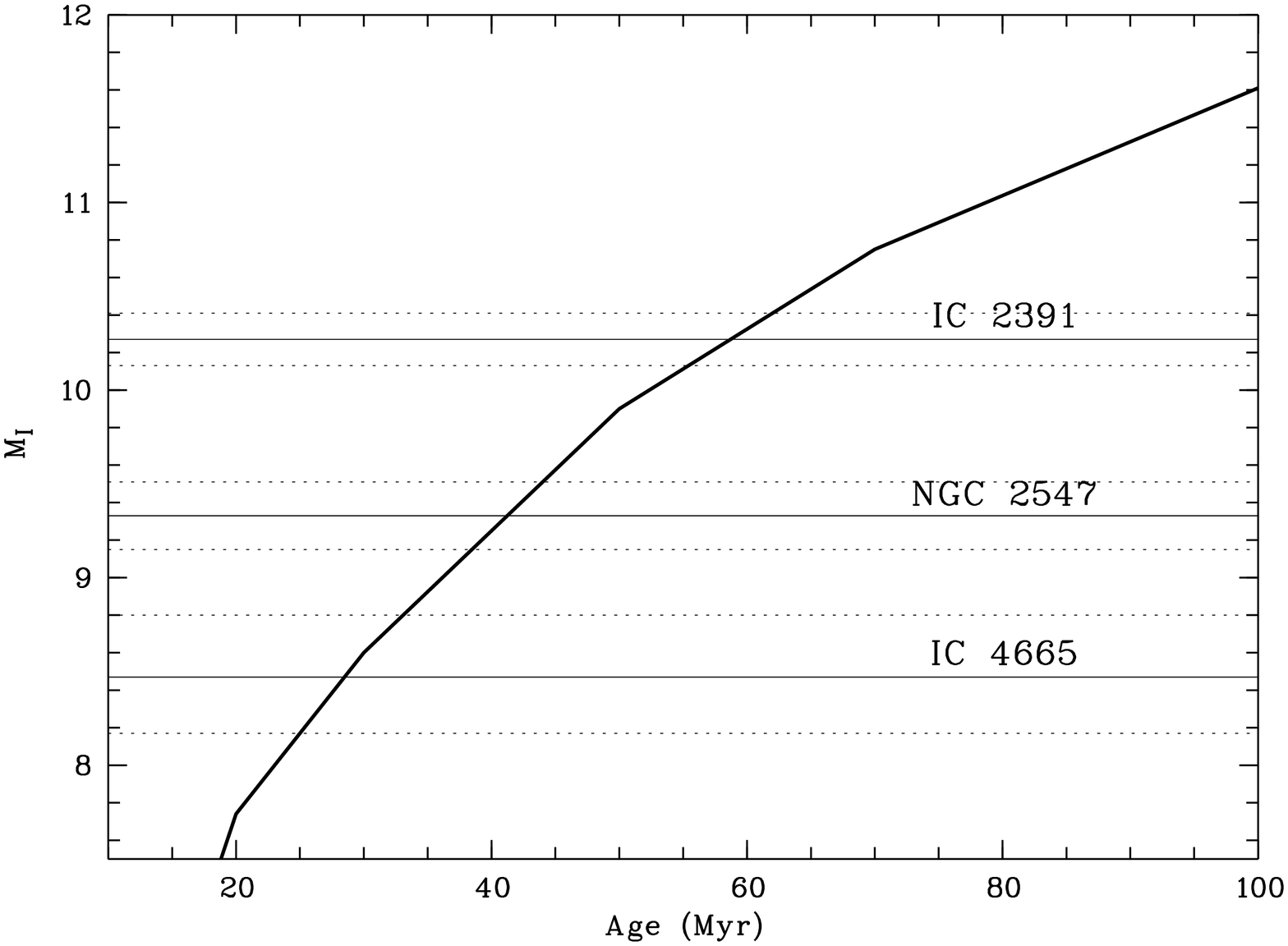}
\includegraphics[angle=-90,width=8.5cm]{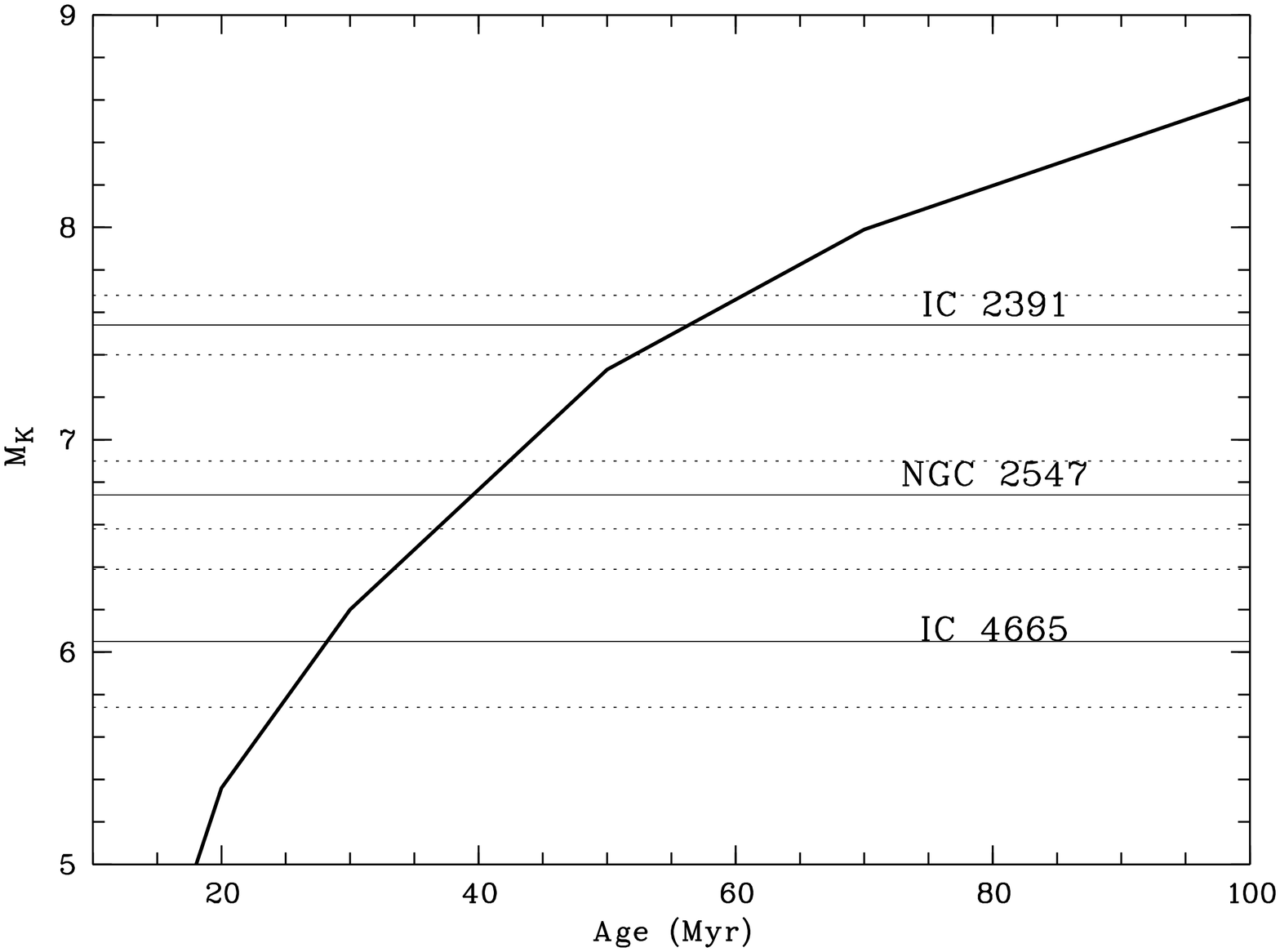}
\caption{Location of currently known LDB of PMS clusters in absolute I$_{c}$
(left) and \ks~(right) magnitudes with uncertainties (dotted lines). 
Clusters are ordered from young to old: IC\,4665, NGC\,2547, IC\,2391. 
The solid curve shows the predictions of the LDB as a function of age
from the evolutionary models of Chabrier \& Baraffe (1997) with $\alpha=1$.  
}    
\label{LDB}
\end{figure*}

The age of IC\,4665 can now be determined using model
calculations. Fig.\,\ref{LDB} displays the
time variation of the absolute I$_{c}$ and K$_{\rm s}$ magnitudes of a star 
which has depleted 99~\% of the
initial Li abundance, according to the models of Chabrier \& Baraffe
(\cite{cb97}). The LDB age that we derive from the two diagrams for
IC~4665 is very similar, namely
t$_{\rm LDB}$(M(I$_{\rm c}))=$28.4$^{+4.2}_{-3.4}$~Myr and 
t$_{\rm LDB}$(M(K$_{\rm s}))=$28.0$^{+4.6}_{-3.8}$~Myr.
Also shown in the figure are the LDBs for NGC~2547 and IC~2391.
In Table~\ref{table3} we summarize LDB magnitudes, ages, and masses for
the three PMS clusters, and using different magnitudes.
Ages and masses were determined using
the models of Chabrier \& Baraffe~(\cite{cb97}); 
LDB magnitudes for IC~2391 and NGC~2547 were taken from
Jeffries \& Oliveira~(\cite{jo05}). M$_{\rm bol}$(LDB) for IC~4665
was computed starting from \ic~and \ks~absolute magnitudes using 
the bolometric corrections (BCs) of Leggett et al. (\cite{legget})
as a function of \vmi~and \imk~colors. Those
corrections were used by Jeffries \& Oliveira~(\cite{jo05}) and we assumed
them for consistency;
we found M$_{\rm bol}$(\ic)$=8.65$ and M$_{\rm bol}$(\ks)$=8.84$.
\setcounter{table}{2}
\begin{table*}
\caption{Properties of the LDB in IC~4665, NGC~2547, and IC~2391.
The upper part lists the absolute magnitudes of the LDB. The bottom
part gives the age and mass of the LDB derived from the three values
of the absolute magnitudes listed in the upper part, 
using the Chabrier \& Baraffe (\cite{cb97}) models with $\alpha=1$.
Errors are the quadratic sum of the uncertainty in distance 
and the uncertainty in the exact magnitude of the LDB.
}
\label{table3}
\begin{tabular}{lc|c|c} \\\hline
   &  {\bf IC~4665} & {\bf NGC~2547} & {\bf IC~2391} \\ \hline
M$_{\rm I_{\rm c}}$ & 8.47$^{+0.33}_{-0.30}$ & 9.33$\pm$0.18 & 10.27$\pm$0.14\\
M$_{\rm K_{\rm s}}$ & 6.05$^{+0.34}_{-0.31}$ & 6.74$\pm$0.16 & 7.54$\pm$0.14\\
M$_{\rm bol}$(\ic)=M$_{\rm bol1}$  & 8.65$^{+0.33}_{-0.30}$ & 9.58$\pm$0.16 & 10.34$\pm$0.14\\ 
M$_{\rm bol}$(\ks)=M$_{\rm bol2}$  & 8.84$^{+0.34}_{-0.31}$ &  & \\  \hline
LDB(M$_{\rm I_{\rm c}}$): t (Myr); M (M$_\odot$) & 28.4$^{+4.2}_{-3.4}$; 0.24$\pm 0.04$ & 35.4$\pm$3.3; 0.17$\pm$0.02 & 49.1$\pm$4.9; 0.12$\pm$0.01\\
LDB(M$_{\rm K_{\rm s}}$): t (Myr); M (M$_\odot$)& 28.0$^{+4.6}_{-3.8}$; 0.24$\pm$0.04 & 34.4$\pm$2.7; 0.17$\pm$0.02 & 50.4$\pm$3.8; 0.12$\pm$0.01  \\
LDB(M$_{\rm bol1}$): t (Myr); M (M$_\odot$) & 26.1$^{+4.1}_{-3.4}$; 0.24$\pm$0.04 & 35.0$\pm$2.2; 0.17$\pm$0.02& 48.8$\pm$3.5; 0.12$\pm$0.01\\ 
LDB(M$_{\rm bol2}$): t (Myr); M (M$_\odot$) & 28.3$^{+4.0}_{-3.5}$; 0.24$\pm$0.04 & & \\ \hline
\end{tabular}
\end{table*}
In the previous section and in Table~\ref{table3}
we considered only the uncertainty due to distance and LDB determination.
However other sources of error are present: {\it i.} Uncertainty
in the \im~to \ic~conversion ($\sim 0.05$ mag), that also reflects into \imk~
colors and thus slightly affects BCs for \ks~magnitudes ($\sim 0.01$~mag); 
{\it ii.} uncertainty
in the \imz~to \vmic~conversion ($\sim 0.3$)~mag, that affects BCs
for \ic~magnitudes (by $\sim 0.1$ mag); {\it iii.}
uncertainty in
reddening, that should not be larger than 0.05~mag;
{\it iv.} finally, the choice
of the bolometric correction vs. color calibration.
By using the calibrations of Bessel~(\cite{bess91}) we
would have obtained
$\sim 0.07$ brighter M$_{\rm bol}$ values. Each of these errors is at most
of the same order of the uncertainty in the LDB determination
and results in an error of about
1~Myr in the LDB age. Remarkably, LDB ages listed in Table~\ref{table3}
are all within 2.4~Myr.
In particular, the LDB age from \ic~magnitude is very similar to that
from \ks~magnitude, that is not affected by the error due to magnitude
conversion. The average of the four values listed in Table~\ref{table3}
is 27.7$\pm 1.1$~Myr.
On top of this, one has to consider the error due to use of
different evolutionary models, which is of the order of 2~Myr, 
as shown by Jeffries \& Oliveira~(\cite{jo05}). Summarizing, our LDB age
for IC~4665 is $27.7^{+4.2}_{-3.5} \pm 1.1 \pm 2$~Myr, where for the error
due to distance and LDB determination we took the average of the errors given
in Table~\ref{table3}.

With an LDB age of 27.7~Myr,
IC~4665 is the youngest of the known PMS clusters.
As to the LDB mass, from Table~\ref{table3} we see that 
for IC\,4465 the Chabrier \& Baraffe (\cite{cb97}) models yield a value of 
M(LDB)$=0.24$~M$_\odot$, independent of the choice of the absolute magnitude.
As expected, this mass is 
above that of NGC~2547 (M(LDB)$=0.17$~M$_\odot$) and IC~2391 
(M(LDB)$=0.12$~M$_\odot$).
\section{Discussion}
\subsection{The age of IC~4665}
As mentioned in Sect.~\ref{intro}, prior to our estimate, Mermilliod
(\cite{merm81}) was the first to determine a nuclear age for IC\,4665
and included it in the age group of  36\,Myr, along
with IC\,2391. Later studies have found
that the cluster could be almost as old as the Pleiades (Prosser~\cite{pros93};
Prosser \& Giampapa~\cite{pg94}), although Prosser~(\cite{pros93}) noted that
the sequence of cluster candidates in the I$_{\rm K}$ vs. \imk~diagram
was suggestive of a rather young age.
Our analysis confirms the young age, making IC~4665 the youngest cluster
for which the LDB has been detected, and, equally important,
allows us to firmly establish its PMS status. In addition, the LDB age 
matches the nuclear age.
This is similar to the case of NGC~2547 (Jeffries \& Oliveira 2005), 
but at variance with IC~2391, $\alpha$~Per, and the Pleiades where the age 
estimates differ by a factor $\sim$1.5.

Although there is no definite explanation for the discrepancy between the TO 
and LDB ages, the difference is usually interpreted as evidence for the
occurrence of some convective core overshooting in high-mass stars
that could lengthen the duration of the main sequence life time. 
The question then arises why a large difference between TO and LDB ages is
instead not found for the two youngest clusters NGC~2547 and IC~4665.
One possibility is indeed that the amount of overshooting in models of massive
stars is a step function of TO mass, with more overshooting needed
for lower TO masses of the older clusters. Still, it is puzzling that IC~2391
and IC~4665 were originally included in the same age group by Mermilliod
(\cite{merm81}), based on the CM and color-color diagrams
for high mass stars. We suggest that a careful re-analysis and comparison
of the two cluster upper
main sequence photometry and CM diagrams should be performed, taking into
account the possible effects of binaries and rotation. 
This, along with
use of updated stellar evolution models including overshooting, might
provide insights on this issue.
\subsection{Isochronal ages and comparison with other clusters}
\begin{figure}
\includegraphics[angle=0,width=9cm]{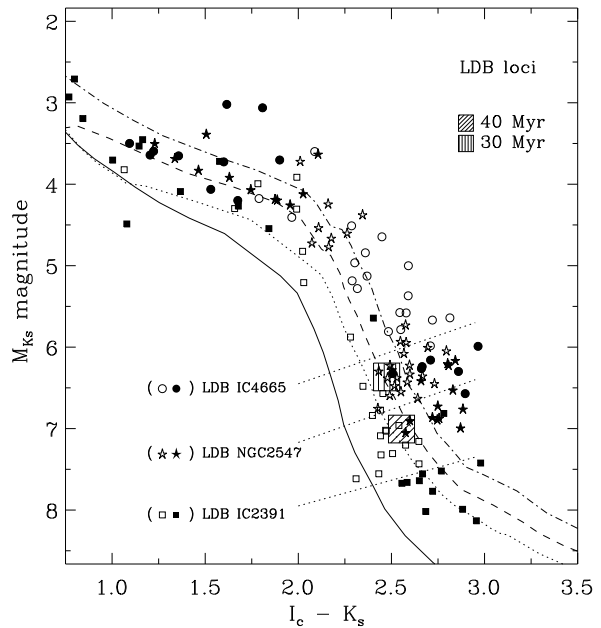}
\caption{The distribution of confirmed members of IC~4665, NGC~2547, and 
IC~2391 in the \ks~vs. \imk~diagram. Filled and empty symbols represent stars 
of each clusters with and without lithium, respectively. The hatched squares 
are the loci of the predicted LDB at different ages, according to the models 
of Baraffe et al. (2002). Also shown are the 20 (dot-dash), 30 (dash), and 
50 Myr (dotted) isochrones from the same models, along with the ZAMS
(solid line). 
The light diagonal curves 
mark the position of the LDB in each cluster, as labeled.}
\label{fig_comp}
\end{figure}

In Fig.~\ref{fig_comp} we compare the distribution of confirmed members of 
IC~4665, IC~2391, and NGC~2547 in the absolute \ks~magnitude
vs. \imk~diagram. The 20, 30, and 
50~Myr isochrones and the predicted location of the LDB for different ages 
from Baraffe et al.~(\cite{bar02}) are also shown, along with the ZAMS
(solid line).

Comparison of isochrones and datapoints up to
\imk~$\sim 2.5$ suggests an age between 20 and 30~Myr
(25$\pm 5$~Myr) for IC~4665 and NGC~2547, in excellent agreement with
the LDB age of the former and slightly younger for the latter. 
At colors \imk~$\ga$2.5 a larger scatter is present in both clusters, 
as well as a discrepancy between the data and the models. This discrepancy 
has already been noted by Jeffries \& Oliveira~(\cite{jo05}) for NGC~2547, 
who found that the \ks~vs. \imk~diagram gives an isochronal age smaller 
than optical diagrams. As a possible explanation, these authors note that
model atmospheres of cool stars do not take into account the
effect of spots, plages, and magnetic activity that could
result in a significant amount of I$-$K excess (see also Stauffer et al.
\cite{stauf03}). Interestingly, the effect of magnetic activity would explain
not only the offset, but also the observed dispersion in the CM diagram.
In this respect, we also mention the recent theoretical study
by Chabrier et al.
(\cite{chabrier}), where an analysis of
the effects of rotation and magnetic fields on the evolution
of M dwarfs is presented. They show that rapid rotation and/or magnetic
field inhibit convection, resulting in a reduced heat flow, and thus
in larger radii and lower effective temperatures than for
normal stars. As a consequence, the stars would appear younger
in a color-magnitude diagram.

As to IC~2391, its sequence generally lies below those of IC~4665
and NGC~2547, yielding an age $\sim 50$~Myr in
good agreement with the LDB age. Thus, we conclude that the overall 
distribution of cluster members shown in Fig.~\ref{fig_comp} indicates a 
smooth progression of increasing ages from IC~4665 to IC~2391.
\section{Conclusions}
We have obtained intermediate-resolution GIRAFFE spectra of 147
cluster candidate low-mass members of IC~4665. The spectra have been
used to measure radial velocities and to establish the presence of the
\ion{Li}{i} 670.8~nm~doublet and H$\alpha$ emission. Using these features
as membership diagnostics, we have identified a subsample of 39 bona-fide
cluster members with a mean radial velocity of $-$15.95$\pm$1.13~\kms.
From the distribution of these stars in
the \im~vs. \imz~and \ks~vs. \imk~color-magnitude diagrams, a clear 
separation of stars with and without Li is found. 
From this boundary, an age of 
$27.7^{+4.2}_{-3.5} \pm 1.1 \pm 2$~Myr is deduced,
making IC~4665 the youngest known PMS cluster with
an LDB determination. The model-dependent mass of the boundary corresponds to
$M_\ast=0.24\pm 0.04$~M$_\odot$. Comparison of the LDB age with 
the standard TO age from Mermilliod~(\cite{merm81}) and that inferred
from isochrone fitting of the cluster low-mass sequence
shows an excellent agreement, a result
similar to that found in NGC~2547 by Jeffries \& Oliveira (2005). This is at
variance with the trend observed in older clusters (e.g., IC~2391, 
$\alpha$~Per, Pleiades) where the LDB age exceeds the TO nuclear age by a factor
of $\sim$1.5. 

The effort to find and characterize PMS clusters in the age range 5--50 Myr 
is being actively pursued by several groups 
and the first results are encouraging. In addition to the two similar 
clusters IC~4665 and NGC~2547 with age 25--35~Myr, two very young clusters 
have been found with ages $\sim$10~Myr, namely NGC~7160 (Sicilia-Aguilar et 
al.~\cite{sici05}) 
and NGC~2169 (Jeffries et al.~\cite{jeff07}), as well as several young moving 
groups in the solar neighborhood (e.g., TW~Hya, $\eta$~Cha, Cha-Near; see 
Zuckerman \& Song~\cite{zuk04}). Now, the next observational challenge is to
fill in the age gap with clusters between $\sim$10 and 30~Myr 
to extend our knowledge on fundamental 
processes related to the early evolution of stars and circumstellar disks,
as well as on star forming process and its duration.

\acknowledgements{We thank the ESO Paranal staff for performing the
service mode observations. We thank the referee, Dr. J. Stauffer, for the
very useful suggestions. We are grateful to Germano Sacco
for providing help with the maximum likelihood analysis of radial velocities.
This work has made extensive use of the
services of WEBDA, ADS, CDS etc. WJDW is grateful for the warm
hospitality and support of the Osservatorio di Arcetri. The research of 
F. Palla and S. Randich has been supported by an INAF grant on 
{\it Young clusters as probes of star formation and early stellar evolution}.}
{}
\clearpage
\onecolumn
\Online
\setcounter{table}{0}
\begin{longtable}{cccccccccccc}
\caption{The 147 candidate members of IC~4665 observed with FLAMES. 
}
\label{table1}\\
\hline
\# & $\alpha_{2000}$ & $\delta_{2000}$ & Ref. & Name & I$_{\rm m}$ & $z$ &
$\rm V_{\rm rad}$ & $\delta\rm V_{\rm rad}$ & Li & H$\alpha$ & Mem. \\
   &   &   & & &  (mag)  & (mag)& (km/s) & (km/s) & & & \\ \hline
\endfirsthead
\caption{continued.}\\
\hline
\# & $\alpha_{2000}$ & $\delta_{2000}$ & R & Name & I$_{\rm m}$ & $z$ &
$\rm V_{\rm rad}$ & $\delta\rm V_{\rm rad}$ & Li & H$\alpha$ & Mem. \\
   &   &   & & & (mag)  & (mag)& (km/s) & (km/s) & & & \\ \hline
\endhead
\hline\endfoot
1&   17 44 45.015&  $+$06 02 11.45& 4& D.08.2.901&    14.691& 14.228&  $-$41.0& 0.8& N& abs& N \\
2&   17 44 47.666&  $+$06 01 52.78& 1& P176&           --- &  --- &     var& ---&N& abs& N \\
3&   17 44 48.120&  $+$06 02 01.22& 3& D.08.2.877&    15.350& 14.731&  $-$29.0& 3.0& N&  em& N \\
4&   17 44 48.354&  $+$05 55 13.39& 3& B.05.30.194&   17.461& 16.764&      var&   ---& ?&  em& ? \\
5&   17 44 51.438&  $+$05 49 53.39& 1& P188&           --- &   ---&  $-$15.0& 3.0& N& abs& N \\
6&   17 44 52.493&  $+$05 48 35.43& 1& P059&          12.614& 12.346&  $-$14.7& 1.0& Y& abs& Y \\
7&   17 44 55.122&  $+$05 49 40.43& 1& P195&          15.763& 15.395&  $-$47.0& 3.0& N& abs& N \\
8&   17 44 57.268&  $+$05 47 20.14& 3& A.00.30.2558&  16.390& 15.74 &  $-$24.0& 4.0& N& abs& N \\
9&  17 44 58.103&  $+$05 51 32.93& 1& P060&    12.623& 12.295&  $-$15.0& 1.0& Y& abs& Y \\
10&  17 45 01.560&  $+$05 56 22.32& 1& P202&          15.529& 15.116&  $-$66.7& 1.4& N& abs& N \\
11&  17 45 01.919&  $+$05 51 55.58& 1& P204&          16.537& 16.102&      SB2& ---& ?&  em& ? \\
12&  17 45 02.388&  $+$06 07 32.45& 4& D.09.2.1536&   12.841& 12.526&  $-$57.1& 2.1& Y& abs& N \\
13&  17 45 03.655&  $+$05 54 33.05& 1& P206&    13.674& 13.319&     13.4& 1.0& N& abs& N \\
14&  17 45 07.463&  $+$05 50 59.39& 1& P064&          13.035& 12.766&  $-$53.6& 0.6& N& abs& N \\
15&  17 45 10.071&  $+$05 49 21.50& 1& P214&          14.124& 13.863&  $-$55.3& 0.4& N& abs& N \\
16&  17 45 10.481&  $+$05 46 17.38& 1& P215&          13.413& 13.154&  $-$14.6& 1.0& Y& abs& Y \\
17&  17 45 10.869&  $+$05 47 57.40& 1& P216&          16.074& 15.614&  $-$61.7& 3.0& N& abs& N \\
18&  17 45 10.906&  $+$05 56 06.81& 1& P217&          15.757& 15.246&  $-$66.0& 2.0& N& abs& N \\
19&  17 45 11.719&  $+$05 45 25.30& 1& P220&          17.183& 16.743&      ---& ---& N& abs& N \\
20&  17 45 12.935&  $+$05 49 50.54& 1& P065&    12.440& 12.17 &      var& ---& Y& abs& Y? \\
21&  17 45 13.301&  $+$05 55 34.93& 1& P222&     13.651& 13.312&  $-$41.7& 1.1& N& abs& N \\
22&  17 45 14.971&  $+$05 49 42.24& 1& P067&          13.912& 13.66 &     57.2& 1.5& N& abs& N \\
23&  17 45 15.190&  $+$06 07 44.77& 3& D.09.30.3076&  16.975& 16.316&  $-$47.0& 4.0& Y&  em& N \\
24&  17 45 16.699&  $+$05 53 56.41& 1& P227&          15.439& 15.012&  $-$11.0& 2.0& N& abs& N \\
25&  17 45 18.453&  $+$05 44 58.84& 1& GPF98-R05&     17.265& 16.639&  $-$7.0 & 3.0& N&  em& N \\
26&  17 45 19.395&  $+$05 47 40.13& 1& P071&          12.696& 12.493&  $-$15.3& 1.3& Y& abs& Y \\
27&  17 45 19.980&  $+$05 46 29.53& 1& P232&          14.147& 13.804&  $-$15.1& 1.2& N&  em& Y \\
28&  17 45 20.486&  $+$05 53 20.74& 1& P233&          16.074& 15.501&  $-$15.8& 2.0& N&  em& Y \\
29&  17 45 20.867&  $+$05 54 27.68& 4& A.00.2.270&    12.761& 12.483&  $-$37.7& 1.5& N& abs& N \\
30&  17 45 22.039&  $+$05 58 25.67& 4& D.09.2.377&    12.823& 12.528&     28.3& 1.5& N& abs& N \\
31&  17 45 23.174&  $+$05 57 06.51& 1& P238&          16.928& 16.314&  $-$14.0& 4.0& ?&  em& Y? \\
32&  17 45 24.851&  $+$06 00 07.91& 3& D.09.30.1207&  16.486& 15.841&  $-$69.0& 4.0& N& abs& N \\ 
33&  17 45 25.078&  $+$05 51 38.75& 1& P075&          12.697& 12.492&  $-$15.8& 1.5& Y& abs& Y \\
34&  17 45 29.077&  $+$05 45 09.24& 3& A.01.30.3322&  17.716& 16.964&      ---& ---& N& abs& N \\
35&  17 45 30.051&  $+$05 48 49.04& 2& P242&          15.139& 14.719&     25.6& 1.0& N& abs& N \\
36&  17 45 30.139&  $+$05 47 07.46& 3& A.01.2.1244&   14.913& 14.405&  $-$15.0& 2.0& N& abs& N \\
37&  17 45 30.454&  $+$05 58 22.28& 4& D.10.2.409&    12.883& 12.595&     22.0& 1.8& N& abs& N \\
38&  17 45 33.091&  $+$05 46 24.35& 1& P155&   12.488& 12.111&  $-$46.0& 1.0& N& abs& N \\
39&  17 45 35.903&  $+$05 49 04.33& 1& P250&          13.513& 13.25 &  $-$68.0& 1.5& N& abs& N \\
40&  17 45 37.654&  $+$05 53 53.40& 1& P155&    12.830& 12.515&     71.7& 1.0& N& abs& N \\
41&  17 45 37.837&  $+$05 45 33.41& 1& P251&          15.201& 14.861&     var& ---& N& abs& N \\
42&  17 45 38.302&  $+$05 44 44.30& 1& P253&          16.226& 15.812&  $-$11.0& 2.0& N& abs& N \\
43&  17 45 39.968&  $+$05 45 16.25& 1& P258&          ---&  ---&      ---& ---& N& abs& N \\
44&  17 45 41.023&  $+$05 54 23.47& 1& P260&          15.378& 15.005&  $-$14.8& 2.0& N& abs& N \\
44&  17 45 42.415&  $+$05 55 41.13& 4& D.10.2.38&     13.177& 12.837&  $-$2.8 & 1.5& N& abs& N \\
46&  17 45 43.521&  $+$05 55 56.46& 1& P262&          17.152& 16.624&     --- & ---& N& abs& N \\
47&  17 45 44.987&  $+$05 49 51.57& 9& LB-3885&       17.059& 16.943&     70.0& 4.0& N& abs& N \\
48&  17 45 56.008&  $+$05 52 45.17& 1& P272&          16.570& 16.069&      SB2& ---& N& abs& N \\
49&  17 45 56.492&  $+$05 48 44.64& 1& K057&           ---  &  ---  &      var& ---& N& abs& N \\
50&  17 45 56.733&  $+$05 52 24.04& 1& P273&          14.505& 14.135&      6.1& 1.0& N& abs& N \\
51&  17 45 59.539&  $+$05 50 45.52& 1& P276&          13.654& 13.395&     16.0& 1.5& Y& abs& N \\
52&  17 45 59.912&  $+$05 36 18.05& 1& P277&          16.098& 15.649&  $-$25.0& 2.0& N& abs& N \\
53&  17 46 01.560&  $+$05 37 11.58& 1& P278&          15.458& 15.101&     10.0& 1.0& N& abs& N \\
54&  17 46 01.604&  $+$05 36 52.79& 2& P279&          15.155& 14.695&     20.0& 1.0& N& abs& N \\
55&  17 46 03.252&  $+$05 33 12.58& 1& P283&          15.474& 14.987&  $-$19.6&   3.0& N&  em& Y? \\
56&  17 46 03.384&  $+$05 50 57.51& 1& P284&          16.435& 15.935&  $-$38.0& 3.0& N&  em& N \\
57&  17 46 03.508&  $+$05 49 42.63& 2& P285&          15.813& 15.335&  $-$47.0& 4.0& N& abs& N \\
58&  17 46 05.764&  $+$05 41 54.62& 2& P286&          15.442& 15.008& $-$110.0& 4.0& N& abs& N \\
59&  17 46 09.712&  $+$05 40 58.13& 1& P290&          13.443& 13.134&  var/SB2& ---& Y& abs/em& Y? \\ 
60&  17 46 10.312&  $+$05 30 56.34& 3& A.08.30.655&   16.940& 16.241&  $-$16.0& 4.0& Y&  em& Y \\
61&  17 46 10.811&  $+$05 42 21.47& 1& P292        &  ---  &   --- &     20.0&10.0& N& abs& N \\
62&  17 46 11.975&  $+$05 41 25.85& 1& P100&          13.074& 12.841&  $-$16.9& 0.6& Y&  em& Y \\
63&  17 46 12.634&  $+$05 38 54.72& 1& P296&          13.940& 13.642&  $-$7.5 & 0.4& N& abs& N \\
64&  17 46 13.813&  $+$05 30 21.42& 2& P298&          16.203& 15.681&  $-$21.0& 4.0& N& abs& N \\
65&  17 46 14.048&  $+$05 36 17.15& 1& P300&          16.916& 16.365&  $-$34.0& 2.0& N& abs& N \\
66&  17 46 15.183&  $+$05 29 25.29& 1& P101&    12.761& 12.359&  $-$72.1& 0.7& N& abs& N \\
67&  17 46 15.234&  $+$05 33 51.83& 1& P303&          15.542& 15.110&  $-$16.0& 2.0& N& abs& N \\
68&  17 46 17.791&  $+$05 45 08.95& 1& P306&          13.344& 13.068&     28.8& 0.5& N& abs& N \\
69&  17 46 18.999&  $+$05 46 20.37& 1& P309&           ---  &    ---&  $-$16.0& 1.0& N&  em& Y \\
70&  17 46 20.068&  $+$05 45 00.19& 1& P311&          16.226& 15.747&  $-$170 &  10& N& abs& N \\
71&  17 46 21.270&  $+$05 29 14.80& 2& P313&   16.736& 15.995&      var& ---& Y&  em& Y? \\
72&  17 46 23.313&  $+$05 37 17.87& 2& P315&   14.876& 14.398&      var& ---& N&  em& Y? \\
73&  17 46 23.892&  $+$05 47 26.94& 1& P317&          14.151& 13.804&      6.4& 0.7& N& abs& N \\
74&  17 46 24.778&  $+$05 35 38.13& 1& P108&          12.718& 12.542&  $-$16.0& 3.0& Y& abs& Y \\
75&  17 46 27.231&  $+$05 29 28.57& 2& P320&          16.096& 15.559&     20.0& 5.0& N&  em& N \\
76&  17 46 28.403&  $+$05 40 18.02& 1& P322&          13.858& 13.598&     35.5& 0.6& N& abs& N \\
77&  17 46 28.923&  $+$05 33 45.03& 1& P323&          15.650& 15.290&  $-$25.4& 0.8& N& abs& N \\
78&  17 46 29.136&  $+$05 31 19.95& 1& P113&          13.421& 13.174&  $-$3.9 & 0.7& N& abs& N \\
79&  17 46 29.480&  $+$05 28 45.69& 1& P326&     13.672& 13.303&  $-$54.1& 0.9& N& abs& N \\
80&  17 46 30.256&  $+$05 30 32.08& 1& P328&          14.353& 14.044&  $-$4.8 & 0.6& N& abs& N \\
81&  17 46 30.476&  $+$05 29 13.98& 1& P329&          15.589& 15.162&      SB2& ---& N& abs& N \\
82&  17 46 31.736&  $+$05 28 35.03& 3& A.09.30.47&    15.940& 15.291&  $-$17.5& 6.0& N&  em& Y \\
83&  17 46 33.977&  $+$05 40 54.06& 1& P331&          13.705& 13.449&  $-$17.4& 0.5& N& abs& Y \\
84&  17 46 34.512&  $+$05 48 53.10& 1& P332&          13.231& 12.979&     10.1& 1.2& N& abs& N \\
85&  17 46 34.731&  $+$05 33 33.56& 2& P333&  16.706& 16.069&  $-$17.0& 3.0& Y&  em& Y \\
86&  17 46 35.156&  $+$05 26 28.85& 1& P334&           ---  &  ---  &  $-$43.0& 2.0& N& abs& N \\
87&  17 46 35.303&  $+$05 36 10.62& 2& P335&          15.902& 15.318&  $-$12.0& 6.0& N&  em& Y? \\
88&  17 46 35.530&  $+$05 31 07.58& 2& P336&          14.910& 14.443&  $-$18.0& 1.0& N&  em& Y \\
89&  17 46 37.712&  $+$05 40 06.38& 1& P337&          16.568& 16.105&      5.0& 1.0& N& abs& N \\
90&  17 46 38.379&  $+$05 35 48.63& 2& P338&          16.685& 16.100&  $-$17.0& 4.0& Y&  em& Y \\
91&  17 46 40.649&  $+$05 28 54.13& 1& P339&     13.822& 13.455&     52.6& 0.6& N& abs& N \\
92&  17 46 40.649&  $+$05 40 19.82& 1& P341&          16.819& 16.329&  $-$43.0& 2.0& N& abs& N \\
93&  17 46 40.811&  $+$05 28 21.36& 3& A.09.30.14&    16.840& 16.118&  $-$13.6& 2.0& ?&  em& Y? \\
94&  17 46 40.906&  $+$05 49 02.67& 1& P344&          16.650& 16.047&  $-$16.0& 2.0& Y?&  em& Y \\
95&  17 46 41.001&  $+$05 44 18.42& 1& P343&          15.228& 14.778&  $-$15.0& 4.0& N&  em& Y \\
96&  17 46 42.209&  $+$05 33 41.98& 1& P346&          13.988& 13.715&  $-$82.8& 0.4& N& abs& N \\
97&  17 46 43.323&  $+$05 35 14.01& 1& P347&          14.768& 14.407&     27.0& 1.0& N& abs& N \\
98&  17 46 43.770&  $+$05 30 07.65& 1& P348&          15.741& 15.177&  $-$17.0& 1.0& N&  em& Y \\
99& 17 46 45.146&  $+$05 26 58.24& 1& P349&          13.306& 13.043&    119.8& 0.7& Y& abs& N \\
100& 17 46 45.359&  $+$05 45 06.06& 2& P350&  16.475& 15.854& $-13.4$ & 3.0& N&  em& Y \\
101& 17 46 46.337&  $+$05 37 09.58& 9& LB3900&        17.270& 17.018&  $-$43.0& 1.0& N& abs& N \\
102& 17 46 46.575&  $+$05 35 06.85& 1& P352&          13.325& 13.108&     43.2& 0.5& N& abs& N \\
103& 17 46 46.736&  $+$05 49 09.35& 1& P119&          12.807& 12.526&  $-$42.9& 0.7& N& abs& N \\
104& 17 46 46.831&  $+$05 47 05.15& 1& P120&          13.251& 12.965&     10.4& 0.6& N& abs& N \\
105& 17 46 47.622&  $+$05 31 38.05& 2& P354&          15.278& 14.800&  $-$16.0& 3.0& N&  em& Y \\
106& 17 46 48.113&  $+$05 47 58.74& 1& P121&             ---&   --- &     84.0& 2.6& N& abs& N \\
107& 17 46 48.743&  $+$05 43 00.62& 1& P122&          13.803& 13.504&   $-$4.8& 2.3& N& abs& N \\
108& 17 46 48.823&  $+$05 41 59.01& 2& P359&          16.276& 15.757&  $-$38.0& 4.0& N& abs& N \\
109& 17 46 48.970&  $+$05 48 54.98& 1& P360&          15.853& 15.451&  $-$35.0& 1.0& N& abs& N \\
110& 17 46 50.237&  $+$05 36 41.06& 1& P362&          15.785& 15.41&   $-$56.0& 3.0& N&  em& N \\
111& 17 46 51.848&  $+$05 31 27.73& 1& P364&          16.027& 15.614&  $-$79.0& 1.0& N& abs& N \\
112& 17 46 52.200&  $+$05 32 36.92& 1& P123&          14.059& 13.780&     49.9& 0.4& N& abs& N \\
113& 17 46 53.013&  $+$05 32 47.02& 1& P366&          14.083& 13.799&  $-$83.3& 0.6& N& abs& N \\
114& 17 46 54.082&  $+$05 31 26.18& 1& P367&          17.898& 17.292&  $-$65.0& 4.0& N& abs& N \\
115& 17 46 54.719&  $+$05 41 59.74& 3& A.10.30.3980&  17.854& 17.108&      var&   --& ?&  em& ? \\
116& 17 46 54.990&  $+$05 37 45.17& 1& P126&          13.465& 13.149&      101& 0.8& N& abs& N \\
117& 17 46 55.159&  $+$05 41 12.70& 2& P369&          16.322& 15.715&  $-$37.0& 3.0& N&  em& N \\
118& 17 46 55.620&  $+$05 30 29.42& 2& P370&          15.412& 14.962&      6.0& 4.0& N& abs& N \\
119& 17 46 56.016&  $+$05 38 34.99& 2& P372&          16.608& 16.028&  $-$18.0& 4.0& Y?&  em& Y \\
120& 17 46 56.418&  $+$05 47 44.62& 1& P374&          14.554& 14.155&  $-$16.2& 0.8& N&  em& Y \\
121& 17 46 56.638&  $+$05 36 07.30& 2& P373&          16.169& 15.537&  $-$16.0& 2.0& N&  em& Y \\
122& 17 46 58.696&  $+$05 47 26.03& 1& P375&          13.377& 13.111&   $-$3.5& 0.8& N& abs& N \\
123& 17 46 59.033&  $+$05 45 44.77& 3& A.04.2.1503&   14.966& 14.422&  $-$51.0& 3.0& N& abs& N \\
124& 17 46 59.319&  $+$05 30 10.30& 1& P129&    12.509& 12.149& $-$116.9& 0.6& N& abs& N \\
125& 17 47 00.227&  $+$05 30 29.04& 1& P377&          13.680& 13.403&  $-$17.0& 2.0& Y& abs& Y \\
126& 17 47 01.787&  $+$05 45 22.94& 1& P379&          16.281& 15.799&  $-$13.0& 2.0& N& abs& N \\
127& 17 47 01.912&  $+$05 30 45.41& 2& P378&          17.119& 16.568&     --- & ---& N& abs& N \\
128& 17 47 03.098&  $+$05 34 46.35& 2& P380&          16.116& 15.544&  $-$17.1& 2.0& N&  em& Y \\
129& 17 47 05.288&  $+$05 43 31.34& 1& P132&          12.882& 12.612&  $-$15.1& 0.6& N& abs& N \\  
130& 17 47 06.467&  $+$05 28 04.14& 1& P382&          13.829& 13.587&   $-$2.6& 0.5& N& abs& N \\
131& 17 47 07.595&  $+$05 36 14.70& 1& P383&          14.467& 14.221&     25.0& 1.0& N& abs& N \\
132& 17 47 08.518&  $+$05 37 37.65& 4& A.10.2.1392&   13.025& 12.744& $-$113.3& 0.8& N& abs& N \\
133& 17 47 09.902&  $+$05 28 13.37& 1& P385&          17.043& 16.116&      ---& ---& N& abs& N \\
134& 17 47 09.961&  $+$05 43 14.06& 2& P387&          15.674& 15.239&  $-$34.0& 1.0& N& abs& N \\
135& 17 47 09.968&  $+$05 29 06.42& 1& P386&    14.712& 14.175&  $-$30.8& 0.7& N& abs& N \\
136& 17 47 11.030&  $+$05 39 55.98& 1& P390&          13.596& 13.324&     61.1& 0.6& N& abs& N \\
137& 17 47 11.660&  $+$05 33 09.98& 3& A.10.30.1418&  17.250& 16.598&    var & ---& ?&  em& ? \\
138& 17 47 11.865&  $+$05 29 24.78& 3& A.10.30.316&   16.236& 15.566&  $-$21.0& 3.0& N&  em& Y? \\
139& 17 47 12.028&  $+$05 43 00.54& 9& IRAS17447$+$054& ---& ---&     SB2?& ---& N& abs& N \\
140& 17 47 12.480&  $+$05 36 33.85& 1& P137&   13.058& 12.713&     67.8& 0.9& N& abs& N \\
141& 17 47 12.532&  $+$05 42 14.95& 1& P394&           ---&     --- &  $-$16.0& 2.0& N& abs& N \\
142& 17 47 16.282&  $+$05 29 49.50& 1& P396&    13.844& 13.457&  $-$16.9& 0.4& N& abs& Y \\
143& 17 47 17.197&  $+$05 31 24.08& 4& A.11.2.358&    12.509& 12.204&     33.7& 0.5& N& abs& N \\
144& 17 47 19.490&  $+$05 44 47.38& 3& A.05.30.3622&  17.248& 16.547&  $-$16.0& 3.0& Y&  em& Y \\
145& 17 47 19.497&  $+$05 30 41.39& 2& P398&    15.373& 14.789&  $-$17.7& 3.0& N&  em& Y \\
146& 17 47 19.636&  $+$05 43 40.74& 1& P399&          16.290& 15.745&  $-$34.0& 3.0& N&  em& N \\
147& 17 47 20.010&  $+$05 46 53.71& 2& P400&          15.269& 14.815&  $-$16.2& 2.4& N&  em& Y \\
\end{longtable}
\end{document}